\documentclass[11pt,twoside]{article}
\usepackage{asp2006}
\usepackage{graphics}
\usepackage{epsf}
\usepackage{lscape}
\def\edcomment#1{\iffalse\marginpar{\raggedright\sl#1\/}\else\relax\fi}
\reversemarginpar

\begin{document}
\title{On the Relationship Between Polar Coronal Jets and Plumes}
\author{Nour-Eddine Raouafi}
\affil{Johns Hopkins University Applied Physics Laboratory, Laurel, MD, USA. e-mail: Nour.Eddine.Raouafi@jhuapl.edu}

\begin{abstract}
We utilize observations from {\emph{Hinode}}/XRT and the Extreme ultraviolet (EUV) imagers onboard {\emph{STEREO}} to study the relationship between coronal jets and plumes. The data were recorded on Apr. 7-8 and Nov. 2-4, 2007. Detailed results are presented for the Apr. campaign along with preliminary analysis of the Nov. observations. We find that $>90\%$ of the identified jets are directly related to plumes (Apr. data). EUV data show that plume haze rose from the same spatial location of more than $70\%$ of the identified jets. The remaining jets occurred in areas where plume material exists already. The jet-plume transition is smooth in some cases and delayed by up to several minutes in others. Short-lived, jet-like events and small transient bright points occur at different locations within the base of pre-existing long-lived plumes. The latter are enhanced after the manifestation of jet-like events. The present observations suggest evidence for X-ray jets as precursors of polar plumes and of their brightness changes.
\end{abstract}

\vspace{-.75cm}
\section{Introduction}

Observations with unprecedented high spatial and temporal resolution from recent space missions, such as {\emph{Hinode}} \citep{Kosugi07} and the {\emph{Solar TErrestrial RElations Observatory}} \citep[{\emph{STEREO:}}][]{Kaiser08}, provide complementary data sets to study the plasma thermodynamic properties within different structures of the solar atmosphere, in particular, coronal jets and plumes. This helps to form a complete picture of these structures.

Coronal jets and plumes were often considered as totally unrelated structures. In fact, their apparent characteristics support this hypothesis. On the one hand, X-ray jets are high-temperature, emissive beams that are collimated by newly open magnetic fields \cite[length: $\sim10^5-10^6$ km; widths: $\sim10^4$ km; see][]{Cirtain07}. The measured plasma outflow speeds within X-ray jets range from $\sim100$ km~s$^{-1}$ to more than $1000$ km~s$^{-1}$. On the other hand, polar coronal plumes appear hazy with no sharp edges. They are significantly cooler, lengthier, and wider than X-ray jets \cite[temperature: $\sim1$~MK; length: $>30~R_{\sun}$; width: $\sim20-40$~Mm; see][]{DeForest97,Wilhelm06}.

It is, however, important to note that coronal jets and plumes share common properties that are fundamental to the physical processes at the origin of their formation and evolution. They are, for instance, both rooted into magnetic flux concentrations of the chromospheric network and are widely believed to form through magnetic reconnection of emergent bipolar flux with the ambient coronal field \citep{Wang98}. Studying the relationship between coronal jets and plumes is important for our understanding of the solar wind and heating of the plasma within the polar coronal holes.

We report on observations of Apr. 7-8, 2007, and also preliminary results from data of Nov. 2-4, 2007, aiming to investigate the relationship between polar coronal jets and plumes.

\section{Observations and Data Analysis}

We utilize data from two observation campaigns of the solar southern and northern polar regions on Apr. 7-8 and Nov. 2-4, 2007, respectively. The {\emph{Hinode}} X-Ray telescope \citep[XRT;][]{Golub07} carried out observations with high spatial (about $1\arcsec$/pixel) and temporal ($<1$ minute) resolution at different time intervals during both campaigns. These data are used to identify X-ray jets. For the present study, we kept only well defined events (several faint and short-lived features were excluded from the sample). For the Apr. campaign, 28 X-ray jets are selected. More events are also identified from the Nov. campaign.

EUV images from the {\emph{STEREO}}/SECCHI/EUVI telescopes are mainly used to identify polar plumes in relation with the X-ray jets selected from XRT data, in addition to several jet events. Data from the 171~{\AA} channel, which is adequate for polar plume emissions, are utilized.

\begin{figure*}[!h]
\begin{center}
\plotone{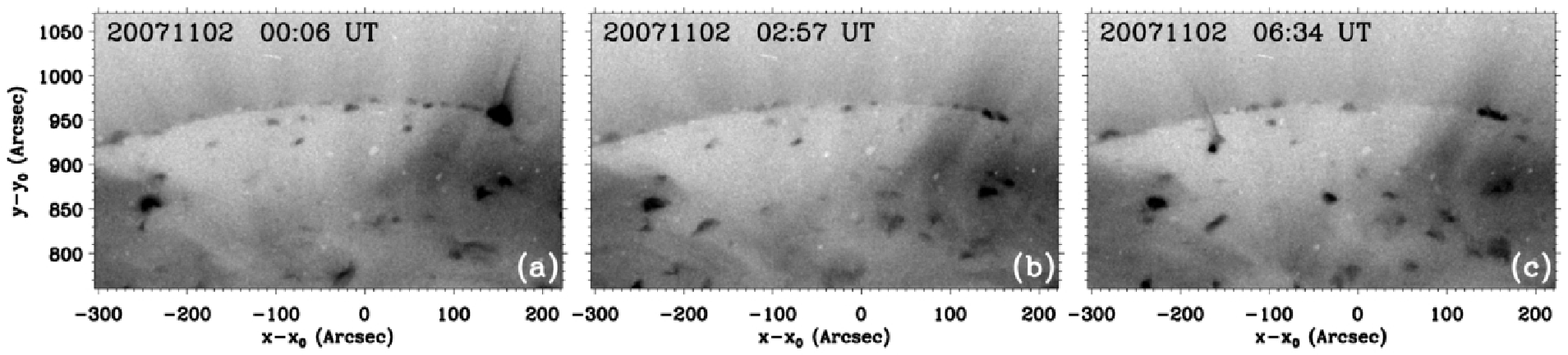}
\plotone{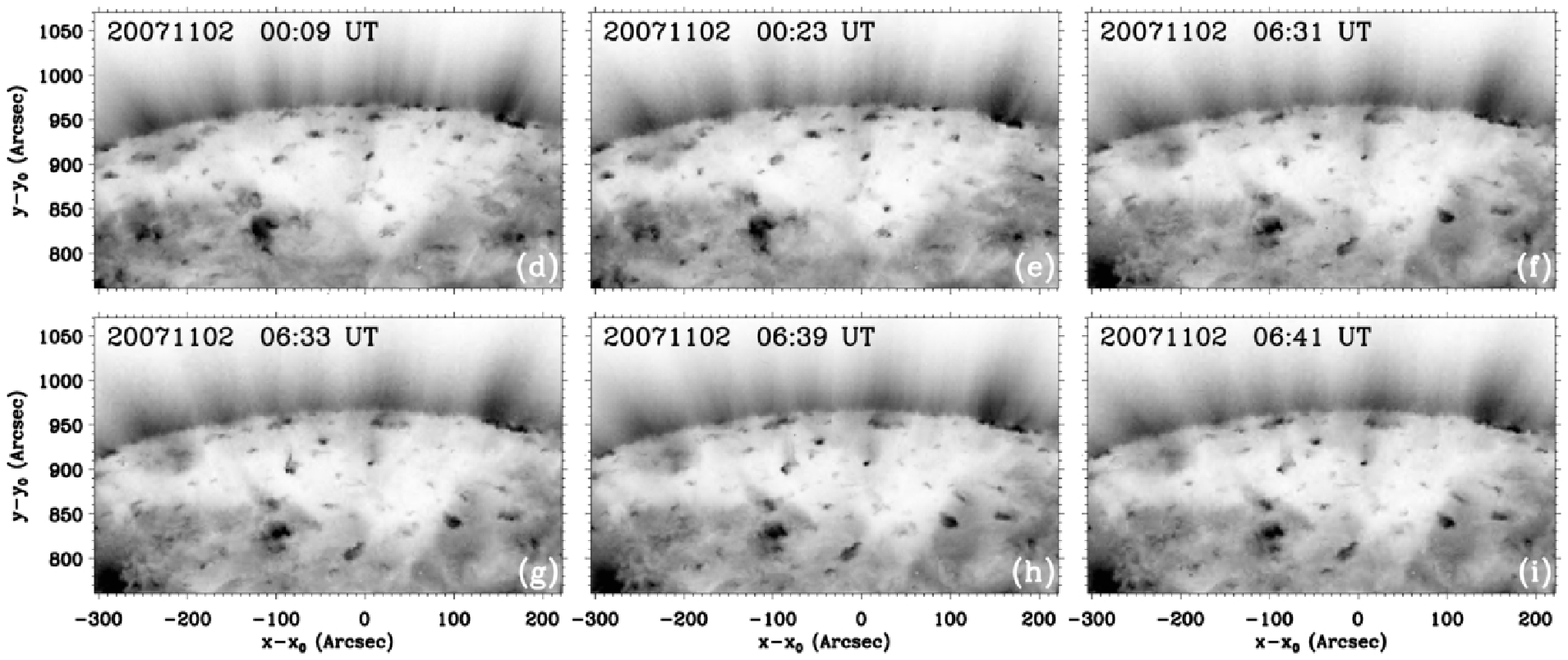}
\caption{(a)-(c): {\emph{Hinode}}/XRT images of the northern polar coronal hole recorded on Nov. 7, 2007, showing two jet events (panel (b) shows the corresponding bright points). (d)-(i): {\emph{STEREO}}/SECCHI/EUVI ÒAÓ images showing polar plume haze clearly rises from the same locations as X-ray jets. \label{fig_xrteuvi}}
\end{center}
\end{figure*}

Fig.~\ref{fig_xrteuvi} illustrates the relation between jets and plumes. Two jets observed by {\emph{Hinode}}/XRT are shown with typical characteristics of X-ray jets (confined-ness, length, and width). The bright point at the base of the event shown in Fig.~\ref{fig_xrteuvi}a is relatively large and lasted for several hours. Several jets erupted from the same location at separate times. The event shown in Fig.~\ref{fig_xrteuvi}c occurred above an initially-weak bright point (see Fig.~\ref{fig_xrteuvi}b) that enhanced just prior to the jet. It fainted after and gradually disappeared.

Although the first event occurred above a bright point very close to the solar limb, which makes off-limb emissions quite diffuse due to integration along the line of sight, it is clear that plume material was well developed after the jet. Taking into account the difference in angular view points between {\emph{Hinode}}/XRT and {\emph{STEREO}}-A ($\sim19^\circ$), the bright point at the base of the jet (and the following plume) is more complex than in the X-ray images. It is also noteworthy that several short-lived, transient jet-like events occurred within the plume. As \citet{Raouafi08} suggested, these may be the powering source of long-lived plumes.

The transition from jet to plume is more evident in the case of the event shown in Fig.~\ref{fig_xrteuvi}c. The jet is very prominent in X-rays and to a less degree in EUV data (Fig.~\ref{fig_xrteuvi}g). The plume haze is, however, conspicuous after the jet disappeared (Fig.~\ref{fig_xrteuvi}h-i). The transition from jet to plume was smooth. The plume lasted for a while after the bright point died out.

Data from the Apr. 7-8, 2007, campaign were studied in more details. We find that about 26 jet events out of 28 were associated with plume material and that polar plumes trailed about 20 of the 26 jets. A number of events showed a time gap between the jet vanishing and the plume appearance. The time gap ranges from minutes to a few tens of minutes.

\begin{figure*}[!t]
\begin{center}
\plotone{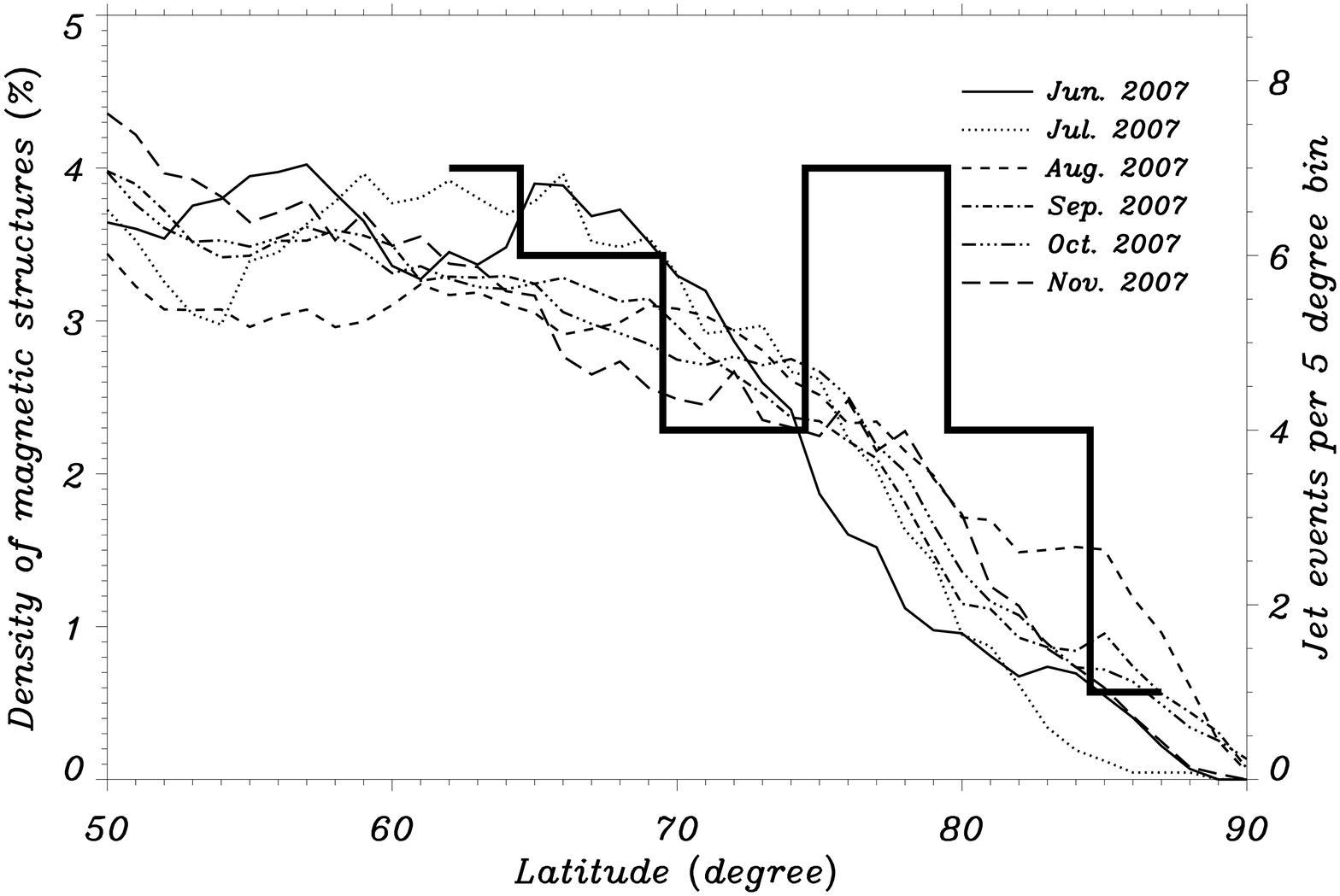}
\caption{Thin lines: monthly normalized (by the observable area surface) distributions of magnetic flux elements as a function of latitude for the north polar cap from June through Nov. 2007 \citep[for details see][]{Raouafi07b}. Thick line: histogram of the 28 jets with a latitude bin size of $5^\circ$. \label{solis_xrt}}
\end{center}
\end{figure*}

\citet{Raouafi07a} studied plasma thermodynamic properties within polar plumes constrained by spectral observations from the {\emph{Ultraviolet Coronagraph Spectrometer}} \citep[{\emph{UVCS}}:][]{Kohl95} onboard the {\emph{Solar and Heliospheric Observatory}} \citep[{\emph{SOHO}}:][]{Domingo95}. They found that in order to reproduce UVCS spectra properties, polar plumes would preferentially be rooted more than $10^\circ$ in latitude away from the solar poles. \citet{Raouafi07b} studied the latitude distribution of polar magnetic field concentrations and found that this distribution decreases significantly toward the solar poles (see Fig.~\ref{solis_xrt}). They concluded that this in agreement with the suggestion that polar plumes are rooted away from the pole.  The histogram of the 28 jets of the Apr. campaign is plotted on the same Figure (thick line). Although the sample is not statistically very large, the histogram shows a trend similar to the latitude polar flux distribution suggesting that more jets (and consequently more plumes) occur away from the solar poles.

\section{Discussion}

The present observations show evidence for a close relationship between polar coronal jets and plumes. We believe that coronal jets are the precursor of polar plume formation, in particular long-lived ones. Jets are likely the impulsive phase of reconnection between newly emergent magnetic flux and the ambient coronal field. Plumes may be the result of slow rate magnetic reconnection and residual heating mechanisms. This is supported by the fact that bright points at the base of plumes usually disintegrate into several parts and also the occurrence of short-lived bright points and transient jet-like events at the base of plumes.

The present results are very important for our understanding of the formation and evolution of fine structures within the polar coronal hole and the physical processes responsible for that. The sample of polar jets and plumes needs to be significantly more important in order to establish a complete picture on the relationship between polar jets and plumes and may be other structures in the solar atmosphere, such as spicules and surges.

\end{document}